# Superconducting elliptical cavities

*J.K. Sekutowicz*
DESY, Hamburg, Germany

**Abstract**
We give a brief overview of the history, state of the art, and future for elliptical superconducting cavities. Principles of the cell shape optimization, criteria for multi-cell structures design, HOM damping schemes and other features are discussed along with examples of superconducting structures for various applications.

## 1    Introduction

There are several milestones worth mentioning that have led to implementation of the superconducting phenomenon in RF acceleration of charged particles. The discovery of superconductivity in 1911 by Heike Kamerlingh Onnes [1, 2], who was the first to liquidize helium in 1908, and the discovery of superconducting properties of niobium (Nb) and lead (Pb) by Walther Meissner [3] in 1928–1934 are certainly some of those.

In parallel, developments in RF acceleration, first proposed in 1924 by Gustaf Ising [4], and the first RF accelerator built by Rolf Wideröe [5] in 1927/1928, were the essential milestones on the particle acceleration technique side for superconducting RF (SRF) technology. In 1961, William Fairbank of the High-Energy Physics Laboratory (HEPL) at Stanford University presented the first proposal for a superconducting accelerator. Three years later the HEPL group of W. Fairbank, A. Schwettman, and P. Wilson accelerated electrons with a lead coated structure for the first time. In 1970, John Turneaure and Ngueyn Viet at HEPL reached a very encouraging result testing at 1.25 K a 8.5 GHz pill-box cavity made of reactor grade niobium [6]. The cavity demonstrated a peak electric (magnetic) field of 70 MV/m (108 mT) on the wall having an intrinsic quality factor[1] $Q_0$ of $8 \cdot 10^9$. The result was proof-of-principle for high-gradient operation of Nb cavities, even though they are made of low RRR material. Finally, the pioneering work of the HEPL group led to the design and construction in the years 1968–1981 of the superconducting accelerator (SCA). The facility became a very successful tool for many years and many experiments, among others for proving the FEL theory of John Madey (1978) in the early 1980s.

The SCA used standing-wave elliptical accelerating cavities. The elliptical shape was invented for suppression of multipacting phenomena in superconducting cavities. The shape in addition allows for less demanding surface preparation, i.e., chemical treatment and high-pressure water rinsing being two inevitable steps to reach high gradients and high intrinsic quality factors. The SCA cavity is displayed in Fig. 1 [7, 1b].

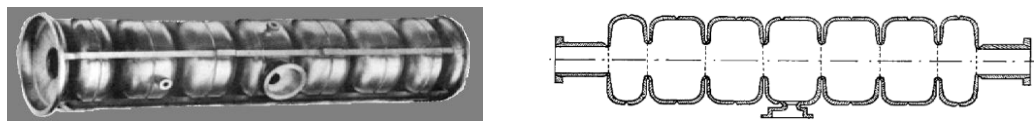

**Fig. 1:** HEPL superconducting cavity: (left) picture of a seven-cell subsection for assembly in a multi-subsection structure; (right) cross-section of the seven-cell structure with attached beam tubes. The opening in the middle of the cell is for the input coupler attachment

---

[1] The quality factor is defined in Section 2.1.

Over the past three decades many superconducting (sc) accelerators have been built, both for high-energy and nuclear physics experiments. Some of them, after years of operation, have already been dismantled, but more are still in operation, under construction, or in a well-advanced R&D phase. Table 1 shows superconducting accelerators, the number of elliptical cavities, length of the SRF installation, and their status. The ultimate goal of the 'superconducting' community is the realization of the International Linear Collider (ILC), since November 2004 the successor of the TESLA project [8]. Figure 2 shows the layout of the 500 GeV option of ILC, with two ~10 km long linear superconducting accelerators. Not much was changed in the TESLA cavity from the time (1992) it was designed [9]. The international R&D effort, since 2004, is mainly to establish a technology ensuring production of ~16 000 TESLA cavities with a high intrinsic quality factor of $10^{10}$ at a nominal gradient of 31.5 MV/m. Recent results of eight TESLA cavities achieving ILC specifications at DESY are summarized in Fig. 3, proving that this demanding specification is achievable in multi-cell structures. The largest facility currently under construction is the European XFEL. The sc linac driving XFEL will be 1 km long. It will be made of 648 TESLA cavities housed in 81 cryomodules. We will discuss TESLA cavities and other alternative cell geometries in more detail later in this lecture.

**Table 1:** Superconducting accelerators with elliptical cavities

| Accelerator | Country | No. cavities | SRF Length [m] | Status |
|---|---|---|---|---|
| TRISTAN | Japan | 32 | 49 | dismantled |
| LEP | Switzerland | 288 | 490 | dismantled |
| HERA | Germany | 16 | 19 | dismantled |
| SCA | USA | 4 | 28 | operational |
| S-DALINAC | Germany | 10 | 10 | operational |
| CESR | USA | 4 | 1.2 | operational |
| CEBAF | USA | 320 | 160 | operational |
| KEK-B | Japan | 8 | 2.4 | operational |
| Taiwan LS | China | 2 | 0.6 | operational |
| Canadian LS | Canada | 2 | 0.6 | operational |
| DIAMOND | UK | 3 | 0.9 | operational |
| SOLEIL | France | 4 | 1.7 | operational |
| FLASH | Germany | 56 | 58 | operational |
| SNS | USA | 81 | 65 | operational |
| JLab-FEL | USA | 24 | 14 | operational |
| LHC | Switzerland | 16 | 6 | operational |
| ELBE | Germany | 6 | 6 | operational |
| CEBAF 12 GeV Upgrade | USA | +80 | 56 | construction |
| SNS-Upgrade | USA | +36 | 33 | design |
| European XFEL | Germany | 648 | 674 | construction |
| ERL Cornell | USA | 310 | 250 | design |
| RHIC Cooling | USA | 1 | 1 | design |
| BEPC II | China | 2 | 0.6 | design |
| X-Ray MIT | USA | Option 176 | Option 184 | R&D |
| Project X | USA | Option 352 | Option 360 | R&D |
| ILC | Option USA | 15 764 | 16 395 | R&D |

CEBAF at TJNAF in Virginia is currently the longest operating SRF installation. Two CEBAF linacs, having 160 superconducting elliptical cavities each, have delivered 6 GeV electrons for nuclear physics experiments over the last 15 years. The upgrade to 12 GeV, which requires 80 additional new-type cavities, operating cw at 19 MV/m, is under construction and first physics experiments at higher energy are scheduled for 2015.

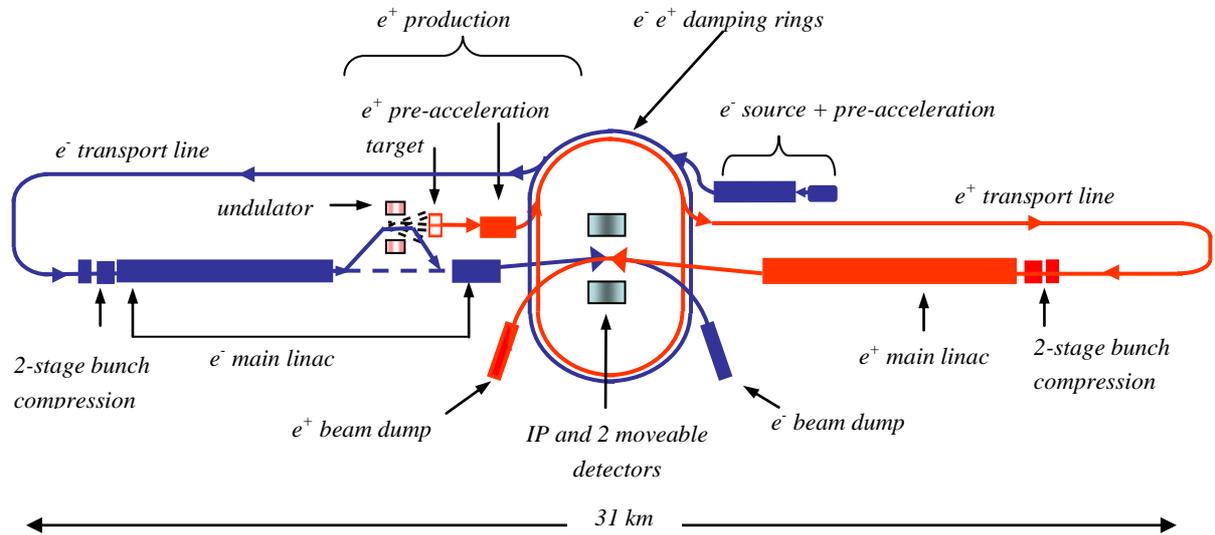

**Fig. 2:** Layout (not to scale) of the ILC for an energy range up to 500 GeV

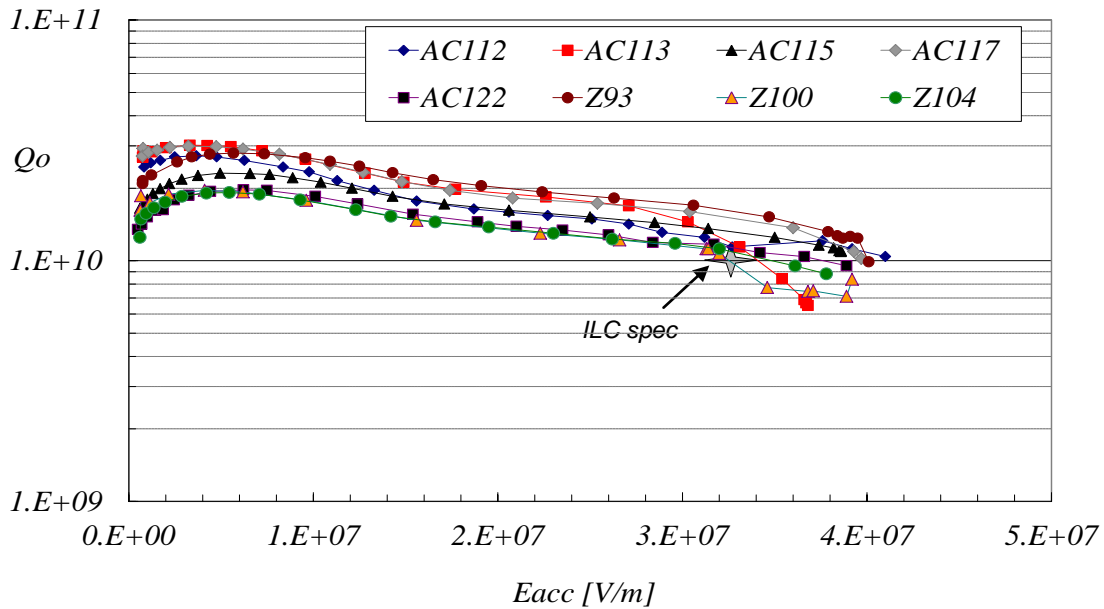

**Fig. 3:** Test results at DESY for eight TESLA cavities achieving the ILC specification

Elliptical cavities are mainly used for acceleration of ultra-relativistic ($\beta = v/c = 1$) particles like electrons and positrons. They can also be used for particles with ß smaller than one. An example of such an application is the superconducting part of the SNS linac accelerating H⁻ ions. In general, RF parameters of an elliptical cavity can be to some extent tailored for a particular accelerator, with respect to the current of accelerated beam, accelerating gradient, cryogenic losses, and mode of operation. In the following two sections, we will introduce RF parameters of an elliptical cavity and discuss criteria for optimization.

## 2 RF parameters

### 2.1 Quality factors

Accelerating structures are microwave resonators (cavities) storing electromagnetic (e-m) energy and sharing it with or receiving it from (ERL case) a charged particle beam traversing their volume. Storing e-m energy always causes its dissipation in the metal the resonator is made of or/and its radiation via openings in the resonator wall. The measures of both processes are intrinsic ($Q_0$) and external ($Q_{ext}$) quality factors respectively. The external quality factors are defined for individual openings. An example of a cavity with volume $V$, metal surface $S$, and three ports is shown in Fig. 4.

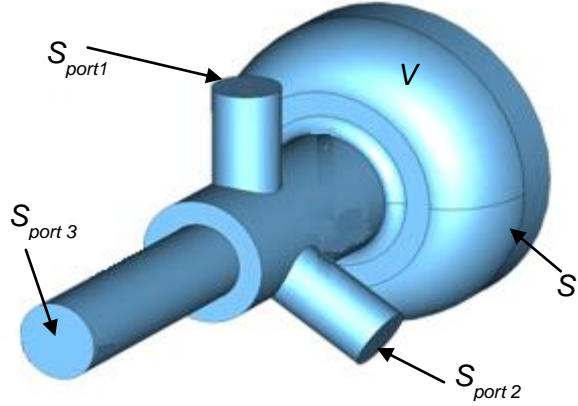

**Fig. 4:** A cavity with openings (ports) coupling in or out stored e-m energy

For each individual resonance, characterized by patterns of the electric field $E$ and magnetic field $H$ in volume $V$ and by resonant angular frequency $\omega = 2\pi f$, one defines the quality factors as follows:

$$Q_0 \equiv \frac{\omega \cdot W}{P_o} = \frac{\omega \cdot W}{\dfrac{R_s}{2} \int_S H^2 ds} \quad , \qquad (2.1)$$

$$Q_{ext} \equiv \frac{\omega \cdot W}{P_{rad}} = \frac{\omega \cdot W}{\dfrac{1}{2} \int_{S_{port}} (E \times H) ds} \quad , \qquad (2.2)$$

where $P_o$ and $P_{rad}$ are dissipated and radiated power respectively, $R_s$ is the surface resistance, and $W$ is the stored energy:

$$W \equiv 2\mu \int_V \frac{H^2}{4} dV = 2\varepsilon \int_V \frac{E^2}{4} dV \quad . \qquad (2.3)$$

Unlike room-temperature cavities for superconducting structures $R_s$,

$$R_s(f,T) = R_{res} + R_{BCS} = R_{res} + 0.0002 \cdot \frac{1}{T} \cdot \left(\frac{f[\text{GHz}]}{1.5}\right)^2 \cdot \exp\left(-\frac{17.67}{T}\right) \qquad (2.4)$$

has two terms: the residual resistance $R_{res}$, practically constant vs. temperature $T$, and the BCS[2] term rapidly increasing with frequency ($\sim f^2$) and decreasing exponentially with $T$. The dependence on $f$ and $T$ has impact on the frequency and operation temperature choice, and will be discussed later with other criteria for a cavity design. The residual resistance is a measure of the surface quality, e.g., purity of the superconductor, roughness, type of oxidation, inclusions on grain boundaries and so on. The general rule is the cleaner the surface the lower the residual resistance.

---

[2] After J. Bardeen, L. Cooper and J. Schrieffer, Nobel laureates for their theory of superconductivity, published in 1957.

## 2.2 Geometric factor

The geometric factor is a ratio of the stored energy and surface integral of $H^2$. Its value equals the intrinsic quality factor $Q_0$ for the unit surface resistance $R_s = 1\ \Omega$, giving a direct comparison for various resonator geometries in respect to $Q_0$ for the same surface resistance. Higher geometric factor means higher intrinsic quality factor and lower energy dissipation for the same surface 'quality'.

$$G \equiv \frac{\omega \cdot W}{\frac{1}{2} \int_S H^2 ds} = \frac{\omega \cdot W \cdot R_s}{P_o} = Q_0 \cdot R_s \quad . \tag{2.5}$$

Again, we will use $G$ as one of the criteria for the minimization of cryogenic load.

## 2.3 Beam–cavity interaction

There are three processes which can take place when a beam of charged particles traverses an accelerating cavity:

- acceleration,
- deceleration (e.g., energy recovery linacs),
- excitation of parasitic modes (Higher or Lower Order Modes).

All three can be described both in time (TD) and in frequency domain (FD).

### 2.3.1 Beam characteristic impedance

Consider a cavity with stored e-m energy $W$. Passing through the interior of a cavity, a charged particle experiences electric force changing its energy (Fig. 5). The change in particle energy $\Delta E_b$ is proportional to voltage $V$:

$$\Delta E_b = q \cdot V \tag{2.6}$$

where $V$ is an integral of the tangential electric field along the particle trajectory:

$$V = \sqrt{\left(\int_{s_a}^{s_b} E \sin(\frac{\omega}{\beta c}(s-s_a))ds\right)^2 + \left(\int_{s_a}^{s_b} E \cos(\frac{\omega}{\beta c}(s-s_a))ds\right)^2} \tag{2.7}$$

and $q$ is the particle charge. Usually, the trajectory is assumed to be a straight line, but in general a curvilinear pass may take place, for example when deflection occurs due to a parasitic dipole mode. Formula (2.7) defines the maximum voltage which can be experienced by a point-like particle at given stored energy. The beam characteristic impedance

$$(R/Q) \equiv \frac{V^2}{\omega W} \tag{2.8}$$

relates the stored energy and maximum accelerating voltage acting on the particle. One should note that *(R/Q)* depends on the cavity geometry, resonant mode field pattern, and on assumed trajectory. It is a measure of how effective the beam–cavity energy exchange is. This effectiveness is higher when *(R/Q)* is larger. *(R/Q)* is the FD parameter.

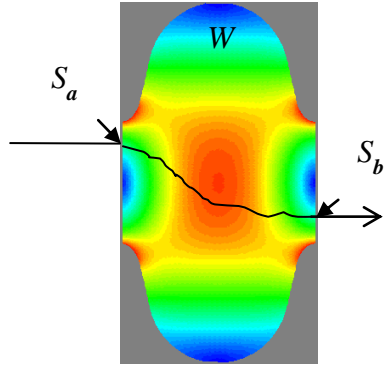

**Fig. 5:** A particle traversing an accelerating cavity along a curvilinear trajectory

### 2.3.2 Longitudinal and transverse loss factors

When a charged particle traverses a cavity free of stored energy it induces energy in all resonant modes, whose projection of the electric field on trajectory is not equal to zero. The deposited energy (wake) is due to the difference in the inducted surface charge density, depending on the distance between the trajectory and the metal wall. The charge density difference causes current flow on the surface, accompanied by magnetic and electric fields. The wake is a superposition of all excited modes. The wake is called longitudinal (transverse) when it is superimposed with monopole (dipole, quadrupole …) modes. The amount of energy lost by charge $q$ is

$$\Delta U_\parallel = k_\parallel \cdot q^2 \quad \text{for monopole modes ,} \tag{2.9}$$

$$\Delta U_\perp = k_\perp \cdot q^2 \quad \text{for non-monopole modes ,} \tag{2.10}$$

where $k_\parallel$ and $k_\perp$ are the loss factors for monopole and transverse modes respectively. For a point-like particle the loss factors, which are time domain parameters, are related to the frequency domain parameter *(R/Q)*:

$$k = \frac{\omega \cdot (R/Q)}{4} \quad . \tag{2.11}$$

The relation (2.11) holds for both longitudinal and transvers loss factors. It demonstrates that the time and frequency domain descriptions are equivalent.

### 2.3.3 Additional RF parameters for accelerating mode

We will now introduce four parameters, which are of practical meaning for the design criteria of accelerating structures.

#### 2.3.3.1 Ratio of peak electric field to accelerating gradient

For stored energy $W_{acc}$, the maximum accelerating gradient $E_{acc}$ is

$$E_{acc} = \frac{\sqrt{\omega_{acc} \cdot W_{acc} \cdot (R/Q)_{acc}}}{l_{active}} \quad , \tag{2.12}$$

where $l_{active}$ is the active length of the accelerating structure. The ratio of the maximum electric field $E_{peak}$ on the metal wall to $E_{acc}$

$$\eta_E = \frac{E_{peak}}{E_{acc}} \tag{2.13}$$

shows the sensitivity of the cavity geometry to electron emission from its metal wall. For TESLA structures $\eta_E = 2$ and in some well performing cavities, $E_{peak}$ is higher than 85 MV/m. At such a high electric field, impurities left on or irregularities in the surface turn to emitters generating electrons, which cause radiation, wall heating, or quenching of the cavity. Often the RF performance limit and slope of the $Q$ vs. $E_{acc}$ curve at high gradients can be attributed to these phenomena. $\eta_E$ was one of the criteria for the TESLA cavity design in 1992. However, with the remarkable evolution in surface preparation methods over the last two decades, higher $\eta_E$ could be accepted in newer geometries of superconducting elliptical cavities, e.g., in the low-loss (LL) shape.

*2.3.3.2 Ratio of peak magnetic field to accelerating gradient*

The critical magnetic flux $B_c$ of a superconductor that the cavity wall is made of, and peak magnetic flux $B_{peak}$ on the wall set the limit for the maximum achievable $E_{acc}$. The ratio

$$\eta_B = \frac{B_{peak}}{E_{acc}} \quad (2.14)$$

depending on the cavity geometry only, is one of the criteria for shape optimization. For niobium (Nb), a metallic type-II superconductor, which is commonly used as bulk material for accelerating cavities, the critical magnetic flux $B_{c1}$, below which the superconductor stays in the Meissner phase, is ~185 mT. The first LL cavity, mentioned above, has been developed for the 12 GeV CEBAF upgrade [10]. For its design, one of the optimization criteria was minimization of $B$ on the wall, leading to less energy dissipation at an operation gradient of 19.5 MV/m. The final LL shape has $\eta_B = 3.74$ mT/(MV/m) and the maximum achievable accelerating gradient for that cavity type, when it is made of Nb, is ~50 MV/m.

*2.3.3.3 Dissipation factor*

For the accelerating mode, one can use the product of the geometric factor $G_{acc}$ and beam characteristic impedance $(R/Q)$

$$\alpha_{dis} = G_{acc} \cdot (R/Q)_{acc} \quad (2.15)$$

as a measure for energy dissipation in the cavity wall. Higher $\alpha_{dis}$ means lower dissipation for the same metal wall quality. The product $\alpha_{dis}$ allows for heat load comparison of various cavity shapes when they are made of the same quality superconductor and when they operate at the same accelerating gradient. From the formulae (2.5) and (2.8), we obtain the expression showing it directly:

$$\frac{P_o}{V_{acc}^2} = \frac{R_s}{G_{acc} \cdot (R/Q)_{acc}} = \frac{R_s}{\alpha_{dis}} \quad . \quad (2.16)$$

*2.3.3.4 Cell-to-cell coupling*

If it is practical from the technical and performance point of view, one should use multi-cell structures to lower the investment costs of an accelerator. Single-cell structures have several technical advantages:

- there is no field flatness problem;
- the input coupler transfers less power[3];
- it is easier to damp HOMs;
- cleaning and preparation are less demanding.

Unfortunately, the advantages do not compensate for the enhanced costs of an accelerator built with single- instead of multi-cell structures. These multi-cell structures have a lower cost per unit length and allow for higher real-estate gradients, which is especially relevant for linear accelerators also for cost reasons.

---
[3] Assuming the same beam current, gradient, and phase.

The cell-to-cell coupling enables energy flow along the structure, for both accelerating mode and HOMs. In the following, we will discuss cell-to-cell coupling for the accelerating mode[4]. Many of the discussed features hold for other resonant modes.

Consider as an example the coupling of two identical, cylindrically symmetric resonators, as shown schematically in Fig. 6(a) for two half-cells of an elliptical cavity. The accelerating mode of either resonator has the same frequency $\omega_0$ and field pattern (shown is the contour of the electric field). After the resonators are coupled [Figs. 6(b) and 6(c)], two resonant modes, slightly different in frequency and field pattern in the coupling region (iris), are formed. The field patterns, depending on whether the coupling plane is the symmetry plane for the magnetic or electric field, have different stored magnetic and electric energies in the iris region. This leads to different frequencies. When there is symmetry for the magnetic field, the resonators oscillate in phase in so-called 0 mode, marked with two '+' signs in Fig. 6(b). In the opposite case, when there is a symmetry plane for the electric field, they oscillate in counter-phase in π mode, marked with '+' and '−' signs in the figure. The 0 mode and π mode have $\omega_0$ and $\omega_\pi$ frequency respectively. When magnetic stored energy in the coupling region is higher than the electric stored energy, $\omega_0$ is higher than $\omega_\pi$. When the coupling region stores more electric than magnetic energy $\omega_0$ is lower than $\omega_\pi$. The former case we call magnetic and the latter one electric coupling respectively. For an ideal lossless case (no beam, no wall dissipation, no radiation out through openings), when the coupled resonators are in steady-state, there is no energy flow across the iris. For either mode in our example, one of the transvers vectors in the Poynting vector equals zero. For 0 mode it is the radial electric field. For π mode it is the angular magnetic field. For a real structure, a linear combination of all modes is needed to enable cell-to-cell energy flow. In the considered example both 0 and π mode must be excited; though the additional mode has much smaller amplitude. We will discuss this phenomenon later in the section on the transient state in accelerating structures.

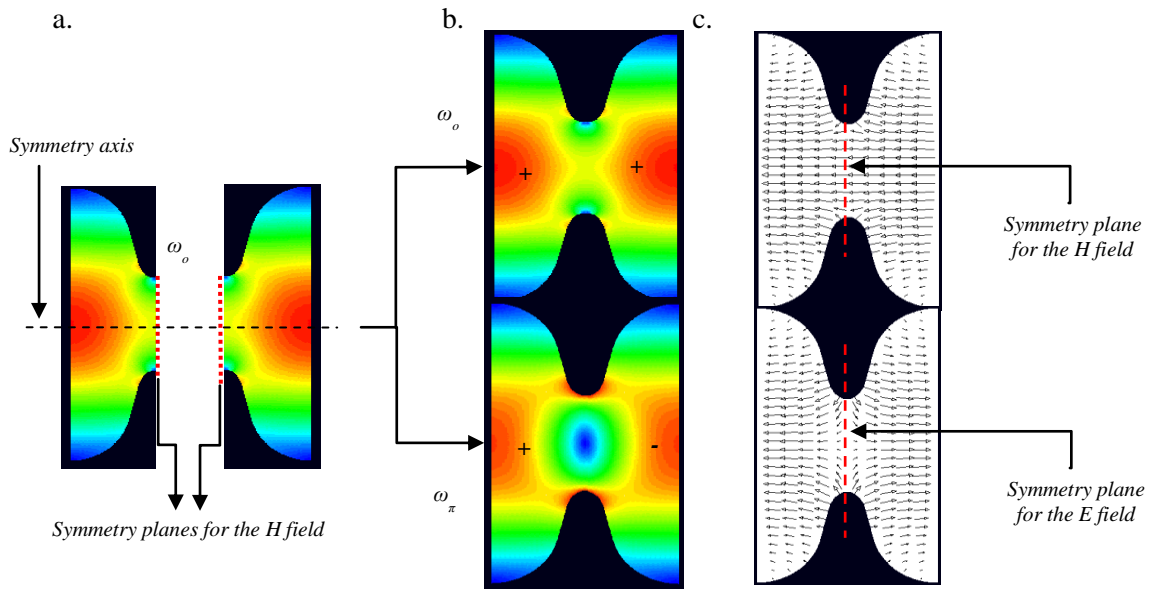

**Fig. 6:** Coupling of two resonators: (a) two identical resonators having the same frequency $\omega_0$ and field pattern before they are coupled, shown is the contour of the electric field for $TM_{011}$ mode; (b) contours of the electric field for two resonances after two resonators have been coupled; (c) electric field lines for two resonant modes

---

[4] For axially symmetric structures, the $TM_{010}$ monopole mode pattern is used for acceleration.

The strength of the cell-to-cell coupling is measured with the coupling factor

$$k_{cc} \equiv \frac{\omega_\pi - \omega_0}{\frac{\omega_\pi + \omega_0}{2}} \qquad (2.17)$$

which is negative for magnetic and positive for electric coupling. For a $N$-cell standing-wave structure, sensitivity of the accelerating field amplitude $E_{acc,i}$ in the $i$-cell to the cell frequency error $\Delta\omega_i$ depends on the coupling strength [11]. When a structure operates in $\pi$ mode, the sensitivity is

$$\frac{\Delta E_{acc,i}}{E_{acc,i}} = a_{ff} \frac{\Delta\omega_i}{\omega_i} \quad , \qquad (2.18)$$

where

$$a_{ff} = \frac{N^2}{k_{cc}} \qquad (2.19)$$

is the field flatness factor, which should be small to keep the structure insensitive to random-cell frequency errors coming from the mechanical tolerances for production, chemical treatment, and cool-down–warm-up cycles.

## 3    Criteria for cavity design

We will limit our discussion to the design criteria for inner cells, because RF properties of these cells are critical for RF properties of the whole accelerating structure.

Figure 7 shows an inner elliptical cell and its shape parameters which one uses to trim the RF properties according to given criteria. We assume for simplicity that the cell is axially symmetric[5]. Not all geometric parameters can be freely chosen. At first, the length of the cell $l_{cell}$ has to be adjusted to the speed of accelerating particles $v$. When particles traverse the cell on the symmetry axis, the maximum energy gain can take place for $l_{cell} = \beta \cdot c/(2f) = v/(2c)$, which provides synchronic acceleration of the beam along the multi-cell structure. Secondly, the radius of the cell $r_{eq}$ must be trimmed to adjust the frequency of the accelerating mode, which is a final step in the design process.

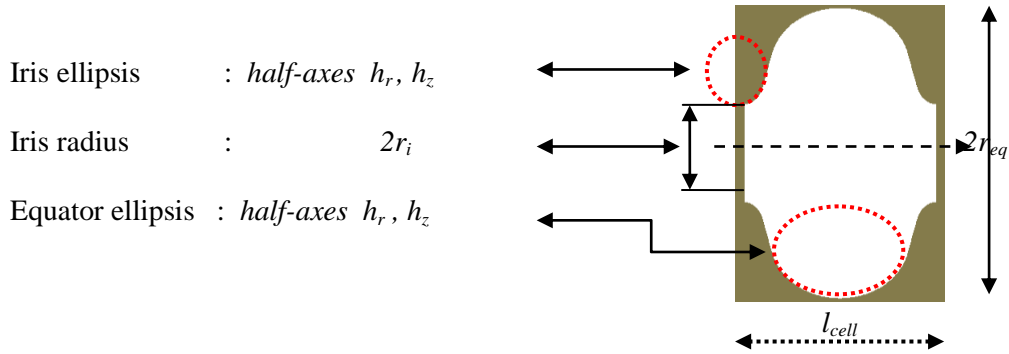

Iris ellipsis     : *half-axes  $h_r, h_z$*

Iris radius       :           $2r_i$

Equator ellipsis  : *half-axes  $h_r, h_z$*

$2r_{eq}$

$l_{cell}$

**Fig. 7:** Inner cell and its shape parameters. The symmetry axis is marked with a dashed line

In the previous section we introduced five RF parameters for the accelerating mode (AM) and two for higher-order modes:

- AM:   *(R/Q), G, $\eta_E$, $\eta_B$, $k_{cc}$*
- HOM:  $k\perp, k_\parallel$

---

[5] In the following, we will use the *(r, φ, z)* coordinate system with the *z*-coordinate parallel to the symmetry axis.

Optimization of these seven parameters, for a given application of a multi-cell accelerating structure, has to be performed with proper adjustment of only five geometric parameters: half-axes $h_r$, $h_z$ of both ellipses, and radius of the iris $r_i$. This obviously leads to conflicts in the optimization process and will require some compromises in the design.

There are three criteria very often used for the inner-cell design, directly related to applications of accelerating structures. Table 2 displays the criteria, RF parameters to be optimized, which geometrical parameters should be trimmed, and examples of existing structures designed for the criteria. Arrows in the table show how both RF parameters and shape parameters should be modified.

**Table 2:** Criteria for the inner-cell design

| Criterion | RF parameter | Improves when | Cavity examples |
|---|---|---|---|
| Operation at high gradient | $\eta_E$ ↓ $\eta_B$ ↑ | $r_i$ ↓ Iris & equator shape | TESLA, HG CEBAF-12 GeV Ichiro /ILC, LL CEBAF-12 GeV |
| Low cryogenic losses | $\alpha_{dis}$ ↑ | $r_i$ ↓ Equator shape | LL CEBAF-12 GeV |
| High $I_{beam}$ ↔ Low HOM impedance | $k_\perp$, $k_\parallel$ ↓ | $r_i$ ↑ | B-Factory, RHIC cooling |

### 3.1  Iris radius adjustment

The third column shows that $r_i$ is a very powerful variable for the inner-cell optimization. When $r_i$ is smaller, $(R/Q)$ is bigger but $\eta_E$ and $\eta_B$ get smaller. An intuitive explanation is that for a smaller iris, electric lines close to the cavity axis are tangential to the particle trajectories on a longer pass and the integral voltage $V$ in formula (2.7) is larger for the same stored energy. The dependence on $r_i$ was studied very carefully in the early 1990s [12]. Results of these studies for the 1.5 GHz elliptic cell summarized in Fig. 8 were the basis for the design of the 1.3 GHz TESLA inner cell.

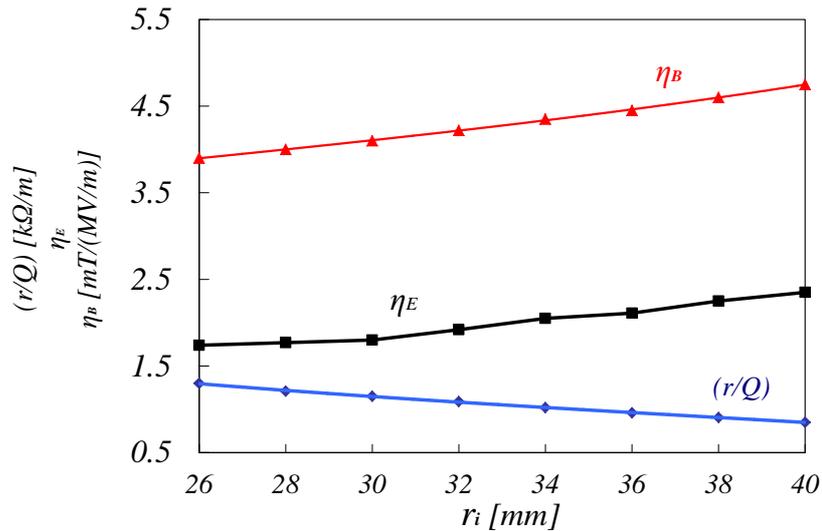

**Fig. 8:** Dependence of $(r/Q)^6$, $\eta_E$, and $\eta_B$ of the accelerating mode on the iris radius

---

[6] $(r/Q)$ denotes $(R/Q)$ per unit length.

Unfortunately, $r_i$ also impacts HOM impedances $k_\perp$, $k_\parallel$, which become larger, and $k_{cc}$, which becomes smaller, for smaller $r_i$. Figures 9 and 10 illustrate dependencies of these parameters on $r_i$ for a 1.5 GHz elliptic inner cell. For the studied range of $r_i$, 20–40 mm, longitudinal and transvers loss factors $k_\perp$, $k_\parallel$ vary by a factor of 2 and 8, respectively. In particular, the change in $k_\perp$ is very fast and needs to be carefully investigated when transvers beam emittance is crucial for the application of an accelerator. Cell-to-cell coupling $k_{cc}$ for the accelerating mode varies even faster vs. $r_i$. For the same $r_i$ range, $k_{cc}$ differs by a factor 10. This is essential for the number of cells one plans to have in a multi-cell structure based on the chosen cell geometry.

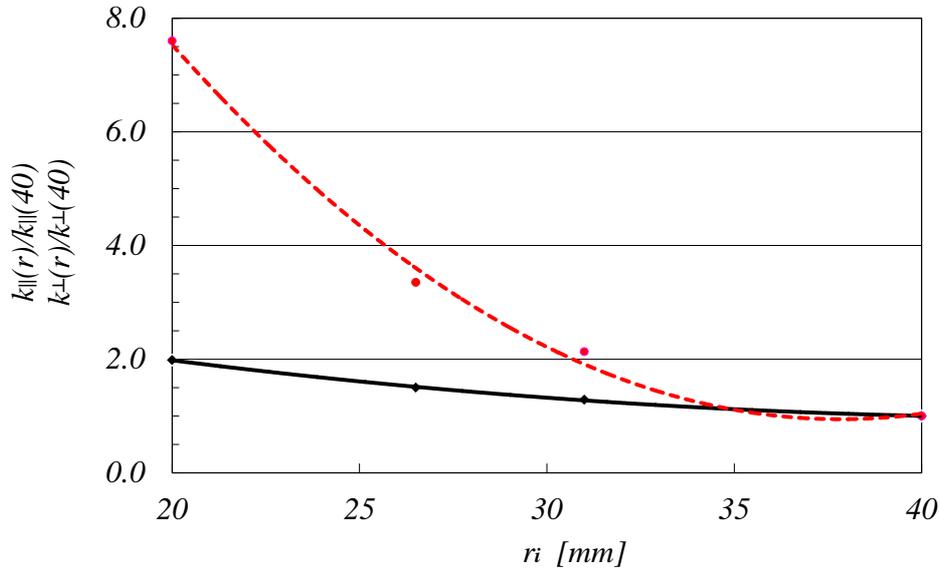

**Fig. 9:** Longitudinal $k_\parallel$ (solid line) and transvers $k_\perp$ (dashed line) loss factors normalized to their values for $r_i = 40$ mm vs. $r_i$, (modelling for 1.5 GHz inner cell)

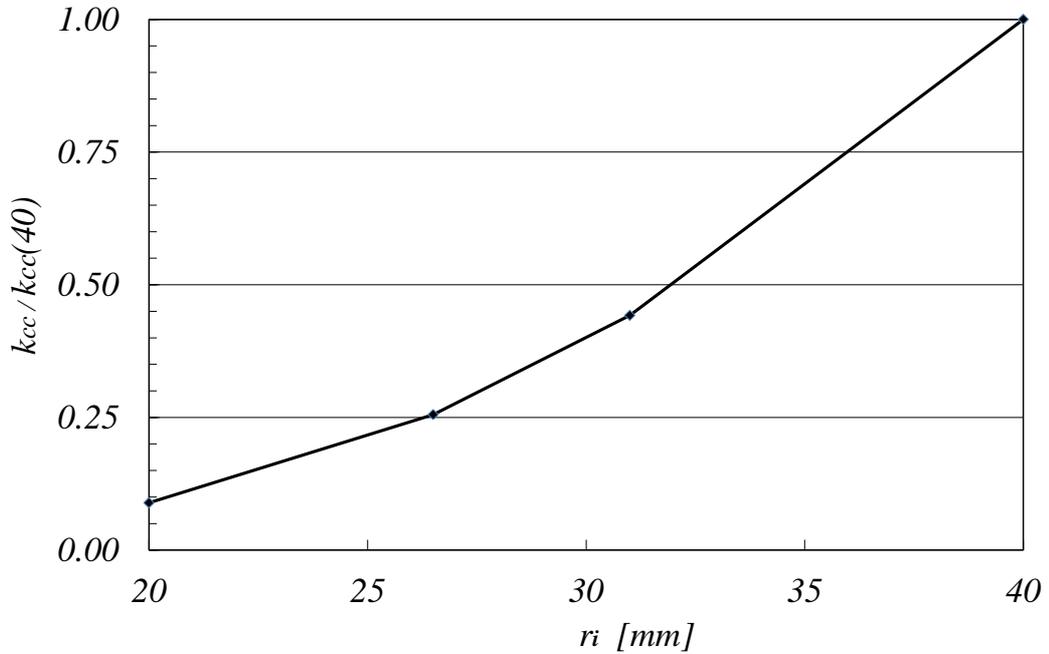

**Fig. 10:** Cell-to-cell coupling $k_{cc}$ for accelerating mode, normalized to its value for $r_i = 40$ mm vs. $r_i$, (modelling for 1.5 GHz inner cell)

## 3.2 Shape of ellipses on iris and equator

In addition to the iris radius $r_i$, $\eta_B$ and $\eta_E$ can be changed with a proper geometry choice for the iris and equator ellipsis respectively. Figure 11 illustrates how the enlarged volume of the equator ellipsis affects the strength of the magnetic flux $B$ on the cavity wall. The cell with more room in the equator region has $(B)^2$ lower by ~16%, which means that power dissipation for that geometry will be lower by this amount. The $B_{peak}$ for this cell is lower by ~8% and thus the maximum achievable gradient is also higher by the same amount than for the other shape shown in the figure. A negative consequence of the enlarged equator ellipsis is that wall slopes are usually almost parallel, making cleaning of multi-cell structures challenging and the cell more sensitive to electromagnetic pressure (Lorentz force detuning). In turn, this enhances the shift of the resonant frequency due to the amount of the stored energy in a structure, which is particularly critical for pulse operating structures at high gradients.

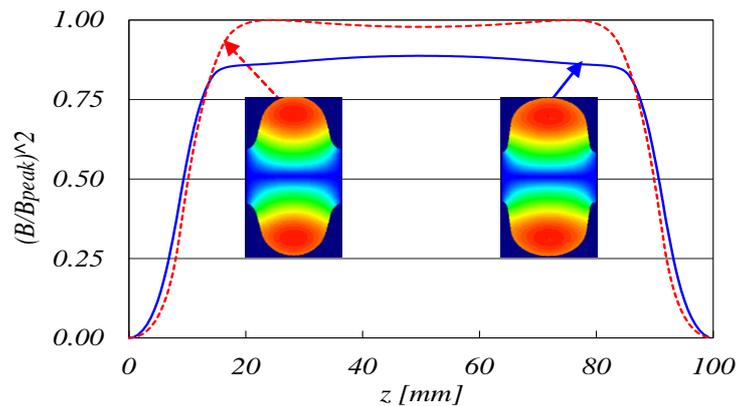

**Fig. 11:** Normalized square of magnetic flux on the metal wall for a cell with less volume (left, dashed line) in the equator region and with larger volume (right, solid line) in the equator region

One can modify $\eta_E$ by changing the aspect ratio of the iris ellipsis. This is illustrated in Fig. 12 for cells with the iris ellipses differing in $h_z$. Both shown cells have the same $f$, $(R/Q)$, and $r_i$. In the example, increased $h_z$ by 20% (5 mm) lowers the peak electric field by ~20% too, making the cell significantly less sensitive to the field electron emission phenomenon. The disadvantage of the longer $h_z$ is a lower cell-to-cell coupling factor $k_{cc}$. Irises are always below cut-off frequency for the accelerating mode, therefore their elongation causes enhanced decay of e-m fields in these regions.

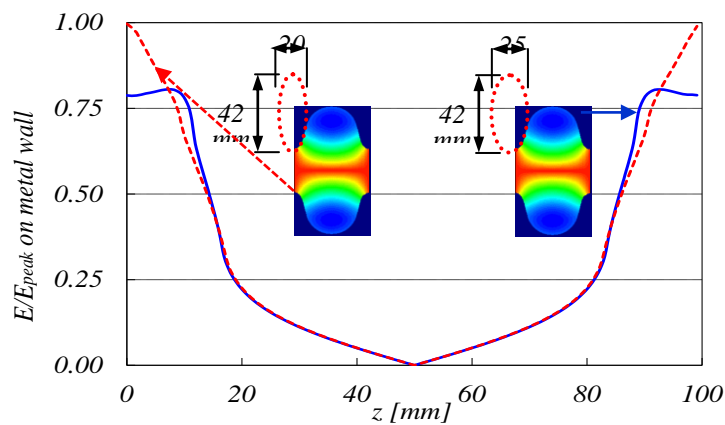

**Fig. 12:** Normalized electric field on the metal wall for cells with shorter half-axis $h_z$ (left, dashed line) and longer $h_z$ (right, solid line) of the iris ellipses

### 3.3 Frequency of the accelerating mode

The choice of resonant frequency for the accelerating mode, for the first two criteria in Table 2, is governed primarily by two contrary requirements. At first, higher frequency makes the characteristic impedance per unit length $(r/Q)_{acc}$ linearly proportionally higher and thus lowers linearly the cryogenic losses per unit length. Formula (2.16) yields

$$P_o = \frac{R_s \cdot V_{acc}^2}{G_{acc} \cdot (r/Q)_{acc} \cdot l_{active}} \quad . \tag{3.1}$$

When $E_{acc}$ and $l_{active}$ are fixed, higher frequency and higher $(r/Q)_{acc}$ seem to be advantageous, but the $R_s$ dependence on frequency [see formula (2.4)] means that overall $P_o$ is proportional to $f$. The fast increase of $R_s$ vs. $f$ can be partially compensated with lower operation temperature $T$ [exponential term in formula (2.4)]. This is why the 1.3 GHz TESLA structures, both in the European XFEL and in the ILC, will have to operate at 2 K (optional 1.8 K), while HERA 500 MHz and LEP 352 MHz structures could operate at a less demanding, from the cryogenic plant point of view, $T = 4.2$ K.

### 3.4 Examples of inner cells and alternatives for the TESLA/ILC shape

Table 3 displays examples of $\beta = 1$ inner cells and their RF parameters. The parameter for which a cell was optimized is underlined. The parameter of the very first CEBAF Original Cornell (OC) cell seems nowadays very conservative, i.e., cell-to-cell coupling, which is very high, even though short 5-cell structures were based on that shape. Almost 20 years later, owing to experience gained in fabrication, chemical treatment and pre-tuning, the LL shape could be proposed for even longer 7-cell 12 GeV upgrade structures having less cell-to-cell coupling. Many fabricated and cold-tested LL structures never demonstrated a problem with field flatness and the best performing cavity reached an $E_{acc}$ of 42 MV/m. An advantage of the LL shape is the enhanced dissipation factor $\alpha_{dis}$, which is 37% higher than the $\alpha_{dis}$ of the OC shape. In the third column, RF parameters of the RHIC electron cooling inner cell are listed. The main criterion for this shape was possibly low HOM impedance because structure had been dedicated to high beam current operation [13]. Both the specification for achievable gradient and the intrinsic quality factor were less critical for that design.

**Table 3:** Examples of inner cells and their RF parameters

|   |   | CEBAF OC (1982) | CEBAF -12 LL (2002) | RHIC Cooler (2003) |
|---|---|---|---|---|
| $f_\pi$ | [MHz] | 1497.0 | 1497.0 | 703.7 |
| $k_{cc}$ | [%] | <u>3.29</u> | 1.49 | 2.94 |
| $\eta_E$ | - | 2.56 | 2.17 | 1.98 |
| $\eta_B$ | [mT/(MV/m)] | 4.56 | 3.74 | 5.78 |
| $(R/Q)_{acc}$ | [Ω] | 96.5 | 128.8 | 80.2 |
| $G_{acc}$ | [Ω] | 273.8 | 280 | 225 |
| $\alpha_{dis}$ | [Ω²] | 26421 | <u>36064</u> | 18045 |
| $k_\perp$ ($\sigma_z = 1$ mm) | [V/pC/cm²] | 0.22 | 0.53 | <u>0.02</u> |
| $k_\parallel$ ($\sigma_z = 1$ mm) | [V/pC] | 1.36 | 1.71 | <u>0.85</u> |

As we mentioned already, the inner cell of the TESLA linear collider structure was designed in the early 1990s. At that time, field electron emission was encountered as a main obstacle in reaching the specified gradient of 25 MV/m. Much effort was made to reduce $\eta_E$ keeping cell-to-cell coupling $k_{cc}$ ~ 2%, since the cell was being designed for a long 9-cell structure. Again, the experience we have gained in fabrication, preparation, and handling allows for revision of the criterion from the 1990s. The maximum achievable gradient for the TESLA inner cell, when it is made of Nb, is 44.5 MV/m. As one can see on Fig. 3, good performing structures approach this limit. When they are properly cleaned, no electron emission is observed during the test. For even higher gradients, one needs to shape the inner cell for lower $\eta_B$. Two geometries of 1.3 GHz inner cell with lower $\eta_B$ were proposed in the years 2002–2004, low loss LL [14] and re-entrant RE [15]. The LL shape with minor modification was implemented later in the Ichiro structures at KEK [16]. All three cells are shown in Fig. 13. Their parameters are displayed in Table 4. The achievable maximum gradient is 52 MV/m and 51 MV/m for the RE and LL shape respectively. LL and RE cells have a higher dissipation factor $\alpha_{dis}$ by ~24% Their cryogenic losses will be lower by the same amount allowing for savings in both investment and the operation costs of cryogenic plant for the ILC. The disadvantages of new shapes are higher HOM loss factors, which, as proven with beam dynamic simulations, are still in an acceptable range for the linear collider.

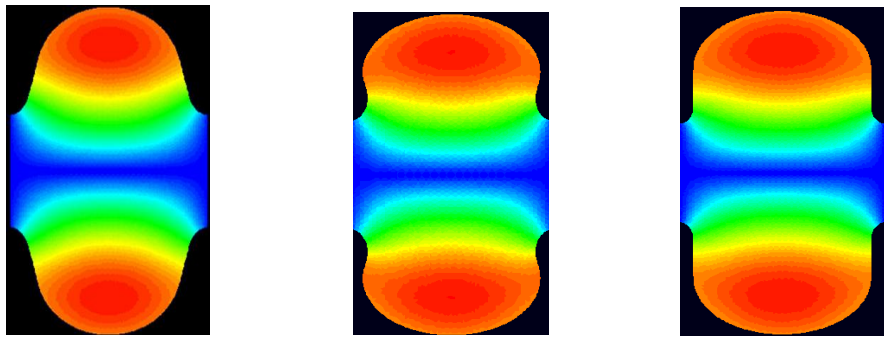

**Fig. 13:** Three inner cells proposed for ILC: original TESLA shape (left), RE shape (middle), LL shape (right)

**Table 4:** Evolution of the TESLA/ILC inner cell and its parameters

|  |  | TESLA (1992) | Re-entrant (RE) (2002/2004) | Low Loss (LL) (2002/2004) |
|---|---|---|---|---|
| $r_i$ | [mm] | 35 | 30 | 30 |
| $k_{cc}$ | [%] | 1.98 | 1.56 | 1.52 |
| $\eta_E$ | - | 1.98 | 2.30 | 2.36 |
| $\eta_B$ | [mT/(MV/m)] | 4.15 | 3.57 | 3.61 |
| $(R/Q)_{acc}$ | [Ω] | 113.8 | 135 | 133.7 |
| $G_{acc}$ | [Ω] | 271 | 284.3 | 284 |
| $\alpha_{dis}$ | [Ω²] | 30840 | 38380 | 37970 |
| $k_\perp$ ($\sigma_z$ = 1 mm) | [V/pC/cm²] | 0.23 | 0.38 | 0.38 |
| $k_\parallel$ ($\sigma_z$ = 1 mm) | [V/pC] | 1.46 | 1.75 | 1.72 |

### 3.5 Additional remarks

When proposing a new shape for sc cavities one needs to investigate whether or not multipacting phenomena can take place and, if so, at what gradient it may happen. Although the elliptic cells were invented to diminish multipacting phenomena, they are not completely free of this process. For example, the TESLA inner cell demonstrates two-sided multipacting close to the equator at 22 MV/m. Fortunately, when the Nb surface is clean, the process is self-destroying and after a certain conditioning time multipacting stops or at least energy dissipated in the process is negligible. This is not always the case, and when designing a cell one should avoid or keep multipacting above the operating gradient level. A very powerful remedy for multipacting is careful cleaning of the Nb surface, which keeps the secondary electron yield low. While cleaning of the Nb surface is crucial for cavity performance, new proposed shapes have to allow for filling the cavity volume with acids, rinsing water, and then for an uncomplicated taking out of all used chemical compounds, without leaving residues on the surface. Establishing a cleaning procedure for single-cell cavities and then adapting the recipe to multi-cell structures is always a long-term process. It took many years for the original TESLA structure, which is the easiest geometry to clean among all three shown in Fig. 13. The LL shape for the CEBAF upgrade, whose wall has more slope than the wall of the LL shape for the ILC, recently reached very good results in the 7-cell structure. Many ILC single-cell cavities of the LL shape, investigated at KEK, reach gradients beyond 45 MV/m. It is still a problem to demonstrate such a good performance for 9-cell structures. A single-cell RE cavity at Cornell University holds the 'world record' of 59 MV/m [17]. The intrinsic $Q$ at this gradient was $3.5 \cdot 10^9$ showing that, with a $B_{peak}$ of 209 mT, Nb was probably already in the mixed state in which $B$ forms vortices that penetrate the superconductor. A 9-cell RE structure has not yet been tested successfully. We should note that such a cavity is very difficult to clean.

## 4  Multi-cell structures and weakly coupled structures

In Section *2.3.3.4* we have listed pros and cons for multi-cell structures. In the following, we will discuss features of a multi-cell, standing-wave, sc structure.

### 4.1  Accelerating mode in multi-cell structure

There are *N* resonant modes of *N* coupled cells, for every resonant field pattern of an uncoupled cell. The modes, which form a so-called pass-band, differ in frequency, field amplitudes in cells, and in cell-to-cell phase advance. The monopole mode, having a $TM_{010}$ field pattern and a π phase advance (π mode) is commonly used for the acceleration of particles[7]. Figure 14 shows the pattern of electric lines for the accelerating mode in an elliptical standing-wave *N*-cell structure, whose properly tuned cells have equal field amplitude along the axis. For acceleration of particles with velocity *v* close to *c* (*β*~1), all cells in a structure usually have the same length $l_{cell} = \beta \cdot \lambda_{acc}/2$, where $\lambda_{acc} = 2\pi c/\omega_{acc}$, for which synchronic acceleration takes place. The active length of a structure is $l_{active} = N \cdot l_{cell}$ and total characteristic impedance is $(R/Q)_{acc} = N \cdot (R/Q)_{cell}$.

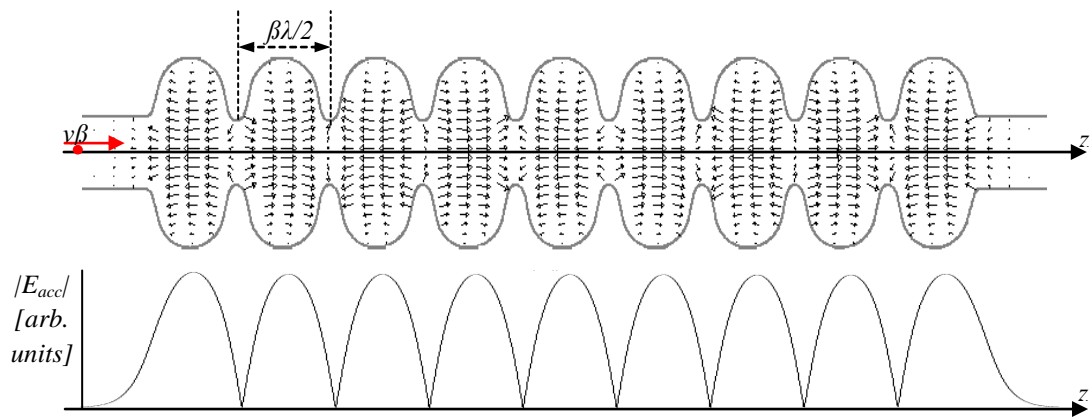

**Fig. 14:** Electric field lines of $TM_{010}$ π mode in an *N*-cell sc elliptical structure (upper), equal amplitudes of the accelerating field on axis for $TM_{010}$ π mode (lower)

---

[7] An exception will be discussed later in the section on weakly coupled structures.

## 4.2 Progress in field flatness preservation

There is always a high probability that, for an unbalanced field profile, the cell with the highest stored energy will limit performance of the whole cavity if the quality of niobium and surface cleanness is comparable in all cells. Thus it is worth while to balance the amplitudes in cells to maximize the accelerating gradient in the cavity. For that, after fabrication, heat treatment, and bulk chemical cleaning, cells are pre-tuned[8] (plastically deformed) to equal the amplitudes. Practically, amplitudes can be adjusted within a few per cent. The last row of Table 5 displays the evolution of $a_{ff}$ over the last 25 years. The increase by a factor of 7 over that time is remarkable. Experience with many TESLA and CEBAF upgrade cavities showed that pre-tuned multi-cell structures, after they underwent final chemical treatment, high-pressure water rinsing, and cold testing, preserved the peak-to-peak field profile with less than 5% difference. Moreover, several ILC Ichiro cavities, in which $a_{ff}$ is the highest at present, have demonstrated that field profile preservation is nowadays possible, and that when designing a new multi-cell cavity one does not need to stay with conservatively low $a_{ff}$.

**Table 5:** Field flatness factor for multi-cell structures

|  | CEBAF OC | CEBAF LL | TESLA | ILC Ichiro |
|---|---|---|---|---|
| year | 1982 | 2002 | 1992 | 2003 |
| $k_{cc}$ | 0.0329 | 0.0149 | 0.0198 | 0.0152 |
| $N$ | 5 | 7 | 9 | 9 |
| $a_{ff}$ | 760 | 3288 | 4091 | 5329 |

## 4.3 HOM excitation and trapping

Excitation of HOM by a multi-bunch beam may lead to dilution of beam quality, i.e., growth of its emittance, particle energy spread in bunch, bunch-to-bunch energy modulation, and to an additional cryogenic heat load. For mitigation of these phenomena, the beam-deposited energy in parasitic resonances has to be coupled out from the cavity and dissipated in external loads. In the following section, the Lumped Element Circuit (LEC) model is used to estimate the beam-deposited energy in a parasitic mode.

### 4.3.1 HOM excitation

Figure 15 shows the generic time structure of a charged particle beam, whose bunches are $\sigma_z$ long[9]. In this example, every single bunch carries the same charge $q$. Bunches are spaced by time $t_b$ within groups. The groups on $N_b$ bunches are separated by time $t_g$. In the frequency domain, the spectrum (Fig. 16) which is a Fourier transform of the beam time structure, has current spectral lines $I_b(f_k)$, depending on $\sigma_z$, $q$, $t_g$, and $t_b$.

A $n$-th parasitic resonance, defined by its angular frequency $\omega_n = 2\pi f_n$, characteristic beam impedance $(R/Q)_n$, and loaded quality factor $Q_{L,n}$ cumulates energy from every spectral line and the total beam-induced power $P_n$ depends on the mode impedance $Z_n(\omega)$ and amplitudes of spectral lines.

Power $P_n$ is estimated by the expression

$$P_n = \frac{1}{2}\sum_k Z_n(\omega_k) \cdot I_k^2 \qquad (4.1)$$

---

[8] The term 'tuned' is usually used for frequency adjustment with a cold-tuner in a cryomodule.
[9] $\sigma_z$ is a standard deviation in Gaussian longitudinal charge distribution.

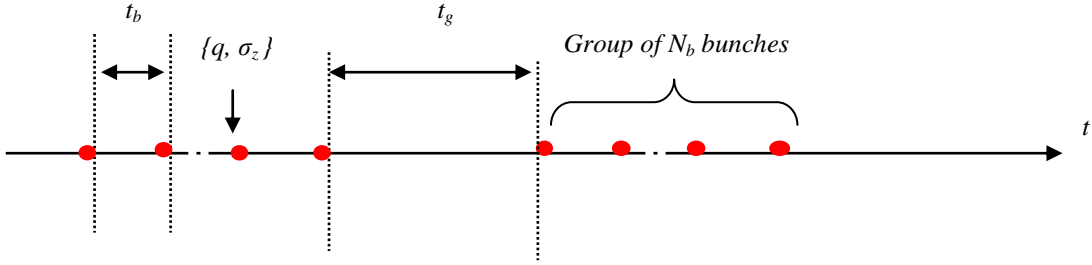

**Fig. 15:** Time structure of an electron beam

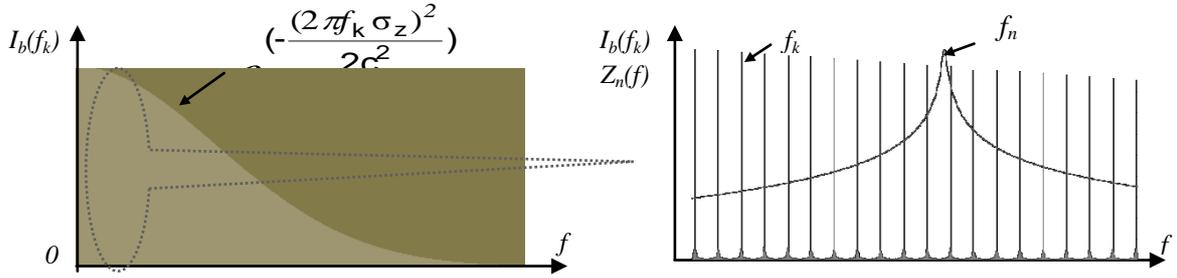

**Fig. 16:** Spectrum of the beam: (left) envelope, (right) zoomed part of the spectrum showing spectral lines overlapping with a parasitic resonance (shown is $Z_n(f)$)

where $k = 0 \ldots \infty$,

$$Z_n(\omega_k) = \frac{(R/Q)_n \cdot Q_{L,n}}{1 + jQ_{L,n}(\frac{\omega_k}{\omega_n} - \frac{\omega_n}{\omega_k})} = \frac{(R/Q)_n \cdot Q_{L,n}}{1 + jQ_{L,n}(\frac{f_k}{f_n} - \frac{f_n}{f_k})} \quad (4.2)$$

is the mode impedance, and

$$Q_{L,n} = (\frac{1}{Q_{0,n}} + \frac{1}{Q_{ext,n}})^{-1} \quad (4.3)$$

is the loaded quality factor of the mode.

There are three ways to minimize $P_n$ and diminish all the above-mentioned phenomena caused by parasitic resonances. At first, $Z_n(\omega)$ can be kept low if $(R/Q)_n$ is low. This remedy applies to non-monopole resonances, whose $(R/Q)_n$ on axis are zero and thus they cannot collect energy from bunches passing a cavity on its cylindrical symmetry axis.

Unlike non-monopoles, the monopole modes demonstrate maximum $(R/Q)_n$ on axis, therefore two other measures are undertaken to keep their fields low. Either method applies both to monopole and non-monopole modes. Thus the second remedy is to keep $f_n$ far from the frequencies of spectral lines $f_k$, which can be done by trimming the cell's geometry. This increases the imaginary part of the denominator in formula (4.2) making the impedance small. The method is effective for non-dense beam spectra and can be helpful to mitigate few HOMs, usually those with very high $(R/Q)$. Finally, extracting a substantial fraction of $P_n$ from the cavity, which makes $Q_{ext,n}$ and therefore $Q_{L,n}$ low, is the most effective way for the suppression of HOMs. For this, structures must be equipped with HOM couplers. These devices are usually complicated and expensive. One should investigate very carefully whether or not they must be implemented because excitation of parasitic resonances is possible and harmful for operation of an accelerator. For superconducting structures, HOM couplers, based on coaxial line or waveguide technique, must be attached to the end of the beam tubes as shown in

Fig. 17. Positioning on cells leads to multipacting and degradation in performance for the accelerating mode, which was observed for the TRISTAN cavities in the early 1980s at KEK. HOM couplers attached to the beam tubes act efficiently on modes having electromagnetic fields at that location. For a multi-cell structure, whose inner cells' HOM frequency can significantly differ from the frequency of end-cells, due to the presence of beam tubes, some modes do not have electromagnetic fields at the HOM coupler locations. The phenomenon is called mode trapping. The trapped modes can be barely suppressed. Their $Q_L$ and $Z$ can be very high, even though the *(R/Q)* is moderate or low. The probability of mode trapping increases with the number of cells. This is one of the limits for making multi-cell structures long when they are designed for high-current operation.

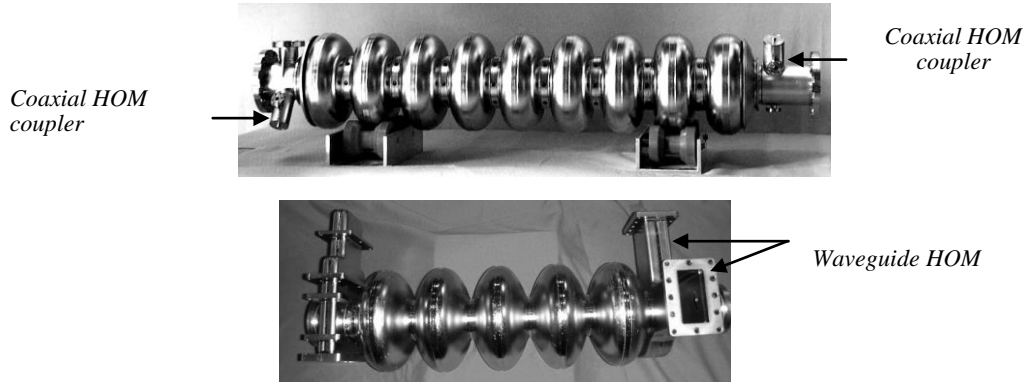

**Fig. 17:** (Top) TESLA structure with coaxial-type HOM couplers attached to the beam tubes, (bottom) CEBAF OC with two HOM waveguide-type couplers

### *4.3.2 Measures against trapping*

HOM trapping has the same nature as the earlier discussed field flatness problem in multi-cell structures. Fewer cells make the HOM field profile less sensitive to HOM frequency spread for individual cells and stronger cell-to-cell coupling makes the field profile more resistant to it. When coupling is small, as for example for the third dipole mode in LEP, HERA, or TESLA structures, or for many quadrupole modes in these structures, even a minor shape deviation may cause substantial changes in the field profile, making damping of modes uncontrollable.

Besides, by making structures shorter one can match the end-cells to fix the field profile for high *(R/Q)* modes if these have small cell-to-cell coupling. In this way the situation becomes less random and HOM couplers can provide sufficient damping. This was done for the third dipole passband of the TESLA structure, whose end-cells have different shapes (asymmetric cavity). One end-cell has been matched to the lower frequency part of that passband and the other one has been matched to the upper frequency part of the passband. Each end-cell has the same accelerating mode frequency as the inner cell. The frequency difference of all three shapes for the first and second dipole passband, which fortunately have strong cell-to-cell couplings, is small and hence their field profiles stay balanced [2b].

Finally, one can split a very long structure into weakly coupled subunits, which are equipped with cold tuners for the field profile balance and with HOM couplers for suppression of parasitic modes in each sub-unit [2b, 18]. Figure 18 shows an example of two 9-cell TESLA-like structures coupled through a $\lambda/2$ long beam tube. For synchronic acceleration of relativistic particles, each subunit operates in π mode and both are in phase (0 mode). Only one fundamental power coupler (FPC) supplies RF power to the whole assembly. This can save half of the couplers and half of the RF power distribution system needed for an accelerator. The savings can be significant when the accelerator is long, as for example the ILC.

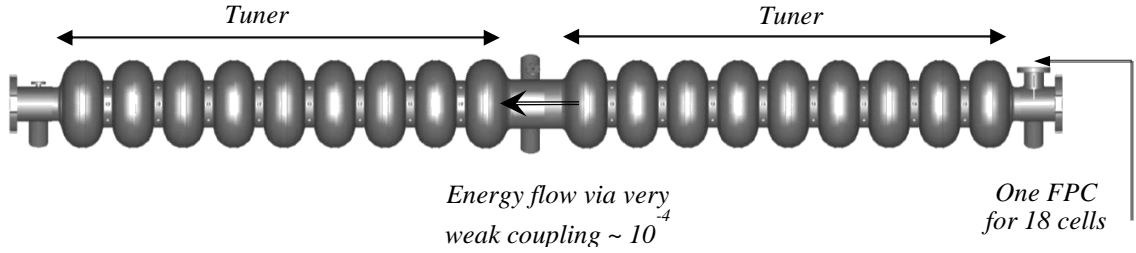

*Fig. 18:* Weakly coupled TESLA-like structures

In addition, this 18-cell 'superstructure' has a good fill-factor. One should note that the field flatness and HOM dumping problem are not enhanced for this assembly since they are fixed for 9-cell subunits. The concept was tested at DESY in 2002. The experiment proved that all specifications, like bunch-to-bunch energy stability, HOM dumping, and field flatness could be achieved for two prototypes made of 7-cell subunits [19].

## 5 LEC

In the design process of multi-cell superconducting structures we use nowadays very advanced computer tools, i.e., 2D-codes (SUPERFISH, SLANS, FEM) and 3D-codes (MWS, HFSS, MAFIA and OMEGA-3P). Even so, a Lumped Element Circuit (LEC) replacement of a multi-cell structure is still very helpful in answering questions regarding the accelerating mode and its passband:

 – How have cells to be tuned after fabrication and chemical treatment to balance the field profile?
 – How does the field profile in cells depend on their frequency errors?
 – How do passband frequencies depend on cell frequency errors?
 – How does the field profile change for non-uniform chemical treatment of cells?
 – What does the transient state (field build-up) in multi-cell structures look like?
 – How stable is the accelerating voltage in multi-cell structures?

The LEC approach is recommended when minor frequency changes, hardly representable in 2D or 3D modelling, are involved in a study. This is the case for the first four questions listed above. Figure 19 shows LEC replacement for the fundamental mode passband of an $N$-cell superconducting structure made of elliptical cells that are capacitively coupled. $L_k$, $R_k$ and $c_k$, which are inductance, resistance, and capacity respectively, can be defined by $f_{acc}$, $(R/Q)_{acc}$, and $Q_{L,o}$ (see caption of the figure). Coupling capacitors $c_{i,\,i+1}$ are defined by the capacity of cells and the cell-to-cell coupling factor $k_{cc}$.

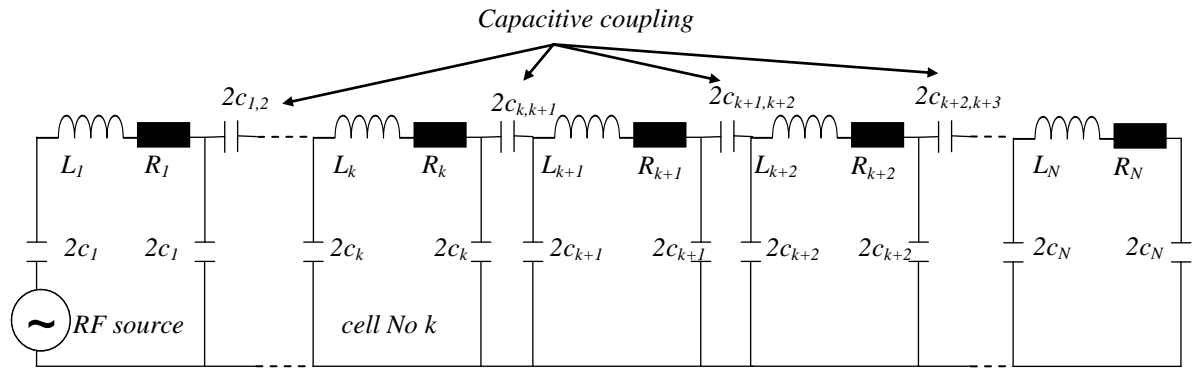

**Fig. 19:** Replacement circuit for an $N$-cell standing-wave structure with the capacitive coupling. The LCR elements for individual cells are defined by the following expressions: $2\pi f_{acc} = (L_k \cdot c_k)^{-0.5}$; $(R/Q)_{acc} = (L_k/c_k)^{0.5}$; $R = (R/Q)_{acc} \, Q_{L,o}$ where $(R/Q)_{acc}$ is the characteristic beam impedance of a single cell, $f_o$ is frequency, and $Q_{L,o}$ is the loaded quality factor of the accelerating mode

## 5.1 Field profile pre-tuning based on LEC

After manufacturing, bulk chemical cleaning, and heat treatment, cells in a multi-cell structure usually have perturbed frequencies and the accelerating mode field profile is unbalanced. There are two consequences of the uneven profile. At first, the cell with the highest field amplitude often limits the maximal operational gradient of a structure, especially when differences in amplitudes are very elevated. Secondly, beam impedances of other modes from the accelerating passband, which for well-tuned structures are usually negligible due to the out-of-phase cancellation, rise and the modes excited by accelerating beam may cause bunch-to-bunch energy modulation. Tuning of cells, which from the RF point of view is an adjustment of their frequencies, is done by plastic deformation of their walls. The difficulty is that one cannot measure directly the frequency of an individual cell as long as it is electro-magnetically coupled with others. The tuning algorithm, if only based on the fact that for capacitive coupled cells lower frequency means lower amplitude[10], can be a very time-consuming procedure requiring many plastic deformation steps and leading to stiffening of niobium, which for each iteration becomes harder to deform. To avoid the lengthy tuning procedures, a new algorithm, in which a frequency correction for every cell is computed from the measured field profiles and frequencies of all passband modes, was proposed in 1985 [20]. The algorithm was first implemented for all HERA structures and then for more than 150 TESLA/FLASH/XFEL cavities tested at DESY. The field profiles and frequencies measured at first, which are eigenvectors and eigenvalues respectively of the coupling matrix for LEC, are then used to compute the frequencies of all cells before tuning. Next, cell-by-cell frequencies of all cells are replaced with the target frequency and then for each adjusted cell, new eigenvalues of the modified coupling matrix are computed. In this way, for example, changes of the $\pi$ mode frequency indicate proper tuning of every single cell. Even though the algorithm requires more RF measurements (frequencies and bead-pulls), it substantially reduces the number of mechanical deformations needed to balance the field profile. For example, for 9-cell TESLA structures having $k_{cc} = 0.02$, reaching 95% field flatness requires that frequencies of individual cells differ by less than ~30 kHz, which corresponds to shape errors less than 2–3 μm, clearly not easy for modelling with 2D or 3D codes.

For such small frequency errors, it is also easier to answer the next three questions with the LEC by means of corresponding analysis.

## 5.2 Transient state and accelerating field stability

The LEC replacement is also sufficient for investigation of the last two questions. By means of the Laplace transform, one can model the time dependence of accelerating fields (voltages) in cells of a multi-cell accelerating structure. Again, as an example, we take the 9-cell TESLA structure. In the studies whose result is shown in Fig. 20, the fundamental mode loaded $Q_{L,o}$ was $3.4 \cdot 10^6$ (matching for the TESLA collider operation). Beginning of the field build-up process shows that there is ca. 80 ns delay between cell No. 1 located close to the input coupler and cell No. 9, which is furthest away from the coupler. The delay does not depend on $Q_{L,o}$ and changes inverse proportionately to $k_{cc}$. At the beam arrival time (right diagram) there is still mode-beating, with a main contribution from the source-supported $\pi$ mode and next lower frequency passband $8\pi/9$ mode. The beating period (distance between every other node) is 1.35 μs, which is the inverse of the difference between frequencies of these two modes. The beating stays for the whole acceleration time, because the energy re-fill in cells for the accelerating mode takes place only if the $8\pi/9$ mode is excited. One should note that the mean voltage (and $E_{acc}$) is much less wavy and therefore bunch-to-bunch energy modulation is not as strong as the cell voltage modulation. One should note that a signal from the pick-up antenna, normally located close to the end-cell, is modulated in the range of $\pm 10^{-3}$.

---

[10] It is opposite for the magnetic coupling for which a lower frequency increases the amplitude.

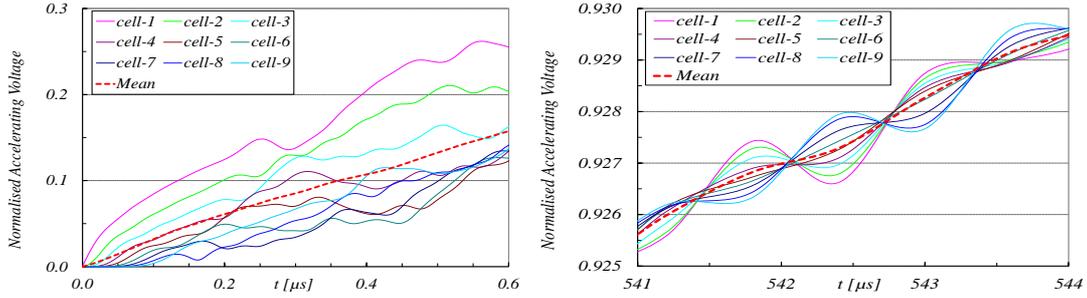

**Fig. 20:** Transient state in the TESLA structure, (left) beginning of the field build-up, (right) voltages at the bunch arrival time

# 6    Production of multi-cell structures and cleaning of the Nb surface

Elliptical half-cells (cups) are shaped in the deep drawing process from niobium sheets ca. 3 mm thick. That fabrication method commonly accepted because of its lower cost compared to turning or milling, does not keep very tight shape tolerances. For this reason cell tuning for the accelerating mode is an indispensible production step and there is always HOM structure-to-structure frequency spread.   Subsequently, pairs of cups are electron-beam welded on irises in dumbbells. The dumbbells together with end-groups (end-cups with beam tubes) are welded on equators in a multi-cell structure. Next, the structures undergo surface preparation (cleaning) procedures, before they are tested in vertical cryostats. The surface preparation has three major steps:

– chemical treatment, buffered chemical polishing (BCP) or electro-polishing (EP),
– high-pressure water rinsing,
– heat treatment.

The first two steps very often are repeated several times, especially when the specified operational gradient is high (> 30 MV/m).  Heat treatment is typically performed once. Buffered chemical polishing is done in a bath of three acids: HF (49%), $HNO_3$ (65%), $H_3PO_4$ (85%), mixed usually 1:1:1 or 1:1:2 by volume, respectively. For electro-polishing one mixes only two acids; 1 part of HF (49%) with 9 parts of $H_2SO_4$ (96%). The anode for the EP process is made of aluminium and the current density is typically 40–70 mA/cm$^2$. For both poly-crystal and large-grain niobium, surfaces after EP are normally smoother than after BCP, however, for large-grain niobium, BCP gives very smooth surfaces too, and for reasons of cost this treatment is preferable.  The sequence for the surface treatments and tuning is listed in Table 6 in more detail [21].

**Table 6:** Sequence of surface preparation procedures and tuning (*courtesy J. Mammosser, TJNAF*)

| Surface preparation procedure | Main purpose(s) |
|---|---|
| Heavy chemical etch (EP or BCP) | Removal of damaged surface layer (100–150 um) caused by fabrication and handling |
| Removal of surface contamination | Ultrasonic cleaning of surface with detergent and deionized water, or alcohol rinse |
| Heat treatment (600–800 °C in vacuum furnace) | Removes hydrogen from the bulk niobium to reduce the risk of Q-disease, release mechanical stress in the material |
| Tuning and mechanical inspection | Field profile correction, check mechanical structure |
| Removal of surface contamination | Ultrasonic cleaning of surface with detergent and deionized water |
| Light chemical etch (EP or BCP) | Remove any risk from damage during handling and furnace contamination |
| High-pressure rinse (ultra-pure water @ 100 Bar) + Class 10 clean room drying of cavity | Reduction of field emission sources, surface particulates |

# 7 High-Q vertical test

## 7.1 Intrinsic quality factor vs. accelerating gradient

After the cleaning procedures discussed above, sc cavities undergo a cryogenic performance test in which the curve $Q_0$ vs. $E_{acc}$ is measured. For that test, a cavity, assembled vertically in a cryostat and immersed in liquid helium for easier cleanness preservation, has an attached input antenna, whose external quality factor $Q_{ext,input}$ is close to its intrinsic quality factor. Very often one uses an adjustable antenna to keep the coupling of the input coupler critical. Unlike the $Q_{ext,input}$, an external quality factor of the pick-up probe $Q_{ext,output}$ is usually much higher than $Q_0$. The schematic test arrangement[11] is shown in Fig. 21.

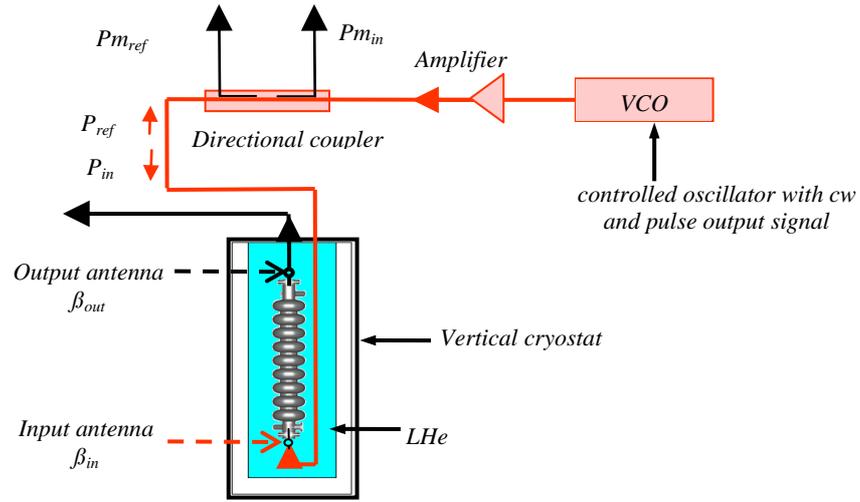

**Fig. 21:** Arrangement for the vertical test: $P_{in}$, $P_{ref}$ are input and reflected power respectively, where $Pm_{in}$, $Pm_{ref}$ are their measured values

The vertical performance test is conducted in three steps. The first two are conducted at low $E_{acc}$ to avoid the influence of phenomena like field electron emission or multipacting. At first, one measures in pulse mode two couplings $\beta_{in} = Q_o/Q_{ext,input}$ and $\beta_{out} = Q_o/Q_{ext,output}$. Here we can employ the LEC replacement of a microwave resonator (see Fig. 22) for analysis of its response to the rectangular RF pulse, whose duration $\tau_p$ is longer than the transient state when the cavity is filled with energy. Two functions given by formulae (7.1) and (7.2) describe the step response, the former for duration of the pulse, the latter when the pulse is off. Figure 23 shows the response for two cases, under critical coupling when $\beta_{in} < 1$ and over critical coupling when $\beta_{in} > 1$. For $\beta_{in} \rightarrow 1$, either case converges to the critical coupling $\beta_{in} = 1$. The calculated and real response signals are depicted with dashed lines. The response signals, if they are measured with a diode and scope (solid lines), do not carry sign information, therefore only their absolute values are displayed on the scope screen.

$$f1(t) = -\frac{1-\beta_{in}}{1+\beta_{in}} - \frac{2\beta_{in}}{1+\beta_{in}} e^{-\frac{\omega_0 t}{2Q_L}} S(t) \quad , \quad \text{for} \quad t \in <0, \tau_p) \tag{7.1}$$

$$f2(t) = f1(t) + f1(t-\tau_p) S(t-\tau_p) \quad , \quad \text{for} \quad t \in <\tau_p, \infty) \tag{7.2}$$

where $S(t)$ is the step function and $Q_L$ is the loaded quality factor.

---

[11] Phase lock loop, keeping VCO at resonant frequency of the tested cavity, is not shown in the figure.

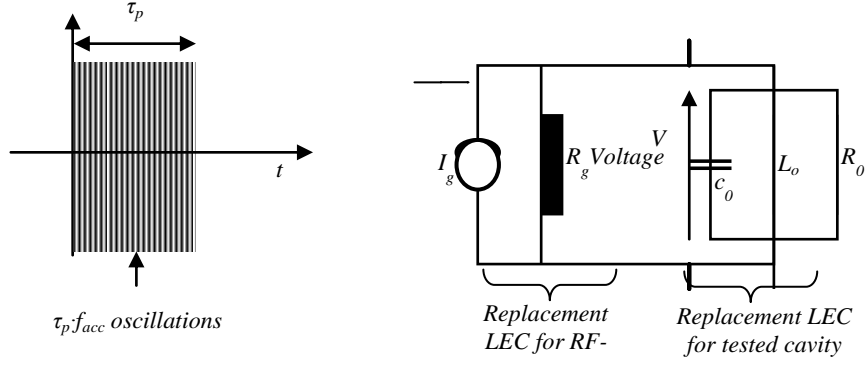

$\tau_p \cdot f_{acc}$ oscillations

**Fig. 22:** Input RF pulse and LEC replacement for the step response analysis

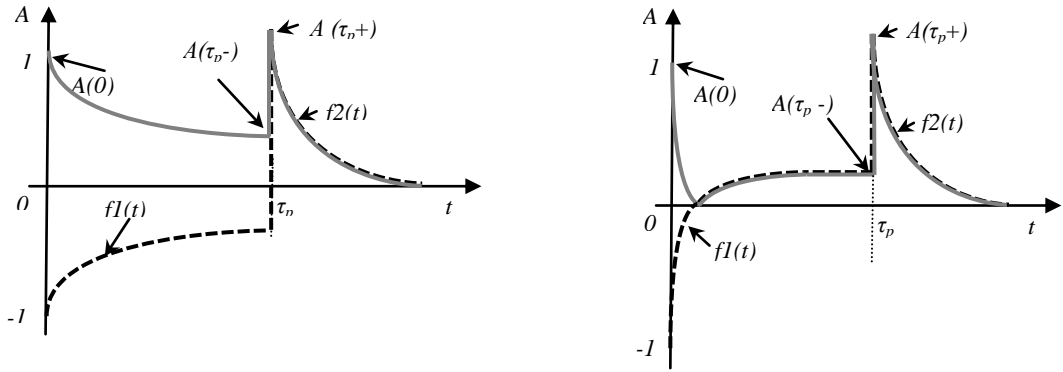

**Fig. 23:** Normalized response signals to the input RF pulse under critical coupling when $\beta_{in} < 1$ (left) and over critical coupling when $\beta_{in} > 1$ (right). In both diagrams, dashed lines indicate real response and analytic signals. The solid lines depict absolute values of the response signals

The coupling $\beta_{in}$ can be computed with one of the three following formulae:

$$\beta_{in} = \frac{A(0) - A(\tau_p-)}{A(0) + A(\tau_p-)} \quad , \tag{7.3.a}$$

$$\beta_{in} = \frac{A(\tau_p+)}{2A(0) - A(\tau_p+)} \quad , \tag{7.3.b}$$

$$\beta_{in} = \frac{A(\tau_p+)}{2A(\tau_p-) + A(\tau_p+)} \quad . \tag{7.3.c}$$

In practice, one computes $\beta_{in}$ with all three expressions and takes the mean value to minimize the measurement errors.

In the second step, one determines $Q_L$ by measuring the decay of the stored energy $W(t)$, right after the pulse is off. The decay is exponential:

$$W(t) = W(\tau_p) e^{-\frac{\omega_0 (t-\tau_p)}{Q_L}} \tag{7.4}$$

where $W(\tau_p)$ is the stored energy at the end of the pulse. On the basis of the measuring signal from the pickup probe at time $t_1$ and $t_2$ one can calculate $Q_L$:

$$Q_L = \frac{\omega_0 (t_2 - t_1)}{\ln(W(t_1)/W(t_2))} \quad . \tag{7.5}$$

When $Q_L$ and $\beta_{in}$ are known and additionally incident power $P_{in}$ and output power $P_{out}$[12] are measured, one can calculate the intrinsic quality factor $Q_0$:

$$Q_0 = Q_L(1+\beta_{in})(1+\frac{P_{out}}{P_{cav}-P_{out}}) \qquad (7.6)$$

where $P_{cav}$ is

$$P_{cav} = \frac{4\beta_{in}}{(1+\beta_{in})^2} P_{in} \quad . \qquad (7.7)$$

$P_{cav}$ can be determined when incident power $P_{in}$ and $\beta_{in}$ are known or when $P_{in}$ and $P_{ref}$ are measured:

$$P_{cav} = P_{in} - P_{ref} \quad . \qquad (7.8)$$

In the final, third step one can continue in the cw mode. The data measured in the previous steps allows for defining the proportionality factor $\kappa$ between $E_{acc}$ and $\sqrt{P_{out}}$:

$$E_{acc} = \frac{1}{l_{active}}\sqrt{(R/Q)Q_0 P_0} = \frac{1}{l_{active}}\sqrt{(R/Q)Q_{ext,output} P_{out}} = \kappa \cdot \sqrt{P_{out}} \qquad (7.9)$$

where

$$\kappa = \frac{1}{l_{active}}\sqrt{(R/Q)Q_{ext,output}} \qquad (7.10)$$

and $P_0 = P_{cav} - P_{out}$ is the dissipated power in the cavity wall. Thus, measuring $P_{in}$, $P_{ref}$ and $P_{out}$ one can calculate the dependence $Q_0$ vs. $E_{acc}$ for the whole test range (a typical diagram is shown in Fig. 3).

## 7.2 Residual resistance test

Residual resistance $R_{res}$ [see formula (2.4)], which is a measure of the surface cleanness and purity of the superconductor, can be estimated as an asymptotic value of the $R_s(f, T)$ when $T \to 0$ K. The test is usually performed at low $E_{acc}$ (ca. 1 MV/m). Measuring the intrinsic quality factor $Q_0$ vs. $T$, by the above discussed methods, we can calculate $R_s$ vs. $T$ [formula (2.5)], if the geometric factor $G$ of the tested structure is known. Figure 24 shows an example of the result.

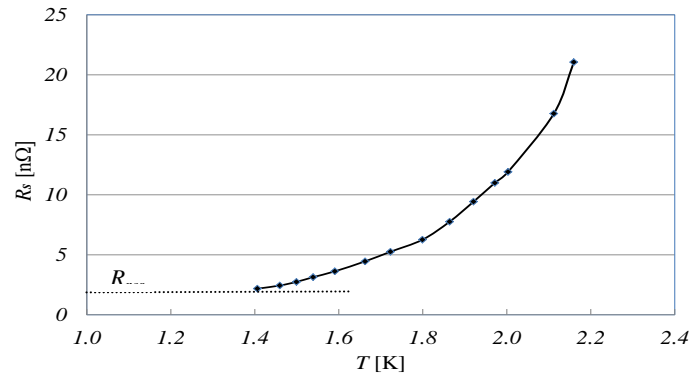

**Fig. 24:** Example of the residual resistance test for a 1.3 GHz cavity *(Courtesy P. Kneisel)* [22]

---

[12] The $P_{out}$ power is very often called transmitted power.

**Acknowledgement**

I express my gratitude to Mrs Katrin Lando for reading this manuscript and many valuable corrections.